\documentclass[aps,prl,twocolumn,showpacs,superscriptaddress,floatfix,preprintnumbers]{revtex4-1}
\pdfoutput=1

\usepackage{graphicx}
\usepackage{bm}
\usepackage{amssymb,amsmath,latexsym}
\usepackage{color}
\definecolor{darkblue}{RGB}{0,0,196}
\definecolor{darkred}{RGB}{196,0,0}
\usepackage[colorlinks=true,linktocpage=true,linkcolor=darkblue,citecolor=darkblue,urlcolor=darkblue]{hyperref}

\newcommand{\p}{\mathbf{p}}

\def\be{\begin{equation}}
\def\ee{\end{equation}}
\def\ba{\begin{eqnarray}}
\def\ea{\end{eqnarray}}

\begin{document}

\preprint{CERN-TH-2020-059}

\title{Non-equilibrium attractor in high-temperature QCD plasmas}

\author{Dekrayat Almaalol} 
\affiliation{Department of Physics, Kent State University, Kent, OH 44242, United States}

\author{Aleksi Kurkela}
\affiliation{Theoretical Physics Department, CERN, 1211 Gen\'{e}ve 23, Switzerland}
\affiliation{Faculty of Science and Technology, University of Stavanger, 4036 Stavanger, Norway}

\author{Michael Strickland} 
\affiliation{Department of Physics, Kent State University, Kent, OH 44242, United States}

\begin{abstract}
We establish the existence of a far-from-equilibrium attractor in weakly-coupled gauge theory undergoing one-dimensional Bjorken expansion.  We demonstrate that the resulting far-from-equilibrium evolution is insensitive to certain features of the initial condition, including both the initial momentum-space anisotropy and initial occupancy.  We find that this insensitivity extends beyond the energy-momentum tensor to the detailed form of the one-particle distribution function. Based on our results, we assess different procedures for reconstructing the full one-particle distribution function from the energy-momentum tensor along the attractor and discuss implications for the freeze-out procedure used in the phenomenological analysis of ultra-relativistic nuclear collisions. 
\end{abstract}

\date{\today}

%\pacs{12.38.Mh, 24.10.Nz, 25.75.Ld, 47.75.+f}


\maketitle

Fluid-dynamic description is a powerful tool in the phenomenological analysis of ultra-relativistic nuclear collisions \cite{Averbeck:2015jja,Jeon:2016uym,Romatschke:2017ejr}. In a fluid-dynamic description of the evolution of the collision system, only a small subset of the degrees of freedom are dynamically evolved. These are quantities derived from the energy-momentum tensor $T^{\mu\nu}$, namely local temperatures, velocities and, in the case of viscous hydrodynamics, information about the shear and bulk viscous tensors. The experiments do not,  however, measure fluid-dynamic variables but rather distributions of particles that have ``frozen out'' and free-stream to the detectors -- the angular and momentum distributions of these particles inform us about the material properties of the fluid created \cite{Teaney:2003kp}. To convert fluid-dynamic fields to particle distributions, a \emph{freeze-out} procedure has to be applied. While the energy-momentum tensor depends only on the first momentum-integral moments of the distribution function, the particle distributions contain information about all the moments. Therefore, the conversion of the fluid-dynamic information to particle distributions necessarily involves injection of new information in the form of theoretical assumptions. The common procedure is to assume that the distribution function has a near-equilibrium form whose deviations from equilibrium arise from formally small corrections -- the shape of the corrections is determined by the response of a linearized collision kernel in some assumed kinetic theory to an infinitesimal strain \cite{Teaney:2009qa,Dusling:2009df}. 

As the freeze-out procedure strongly affects the phenomenological analysis and conclusions about the matter created in ultra-relativistic heavy-ion collisions, it is of great interest to scrutinize quantitatively how well justified are the theoretical assumptions about the shape of the non-equilibrium distribution functions. The need for such scrutiny becomes increasingly important in the case of small systems, \emph{e.g.}, peripheral nucleus-nucleus collisions, proton-nucleus, and high-multiplicity proton-proton collisions where fluid-dynamical description is being applied to situations which most likely remain far from equilibrium throughout their dynamical evolution.

There has been a large body of work quantifying to what extent various different formulations of viscous fluid dynamics are able to reproduce the time evolution of the components of the energy-momentum tensor undergoing expansion in  various geometries. In particular, it has been observed that the hydrodynamic constitutive equations that relate the stress tensor to gradients of the flow fields are well satisfied in systems which are still far from equilibrium at least in systems characterized by flow with a large degree of symmetry \cite{Chesler:2009cy,Kurkela:2015qoa} (for systems with less symmetry \emph{cf.}  \cite{Chesler:2016ceu,Kurkela:2018vqr,Kurkela:2018qeb,Kurkela:2019kip}) -- a feat dubbed \emph{hydrodynamization} without thermalization \cite{Heller:2015dha,Keegan:2015avk,Heller:2016rtz,Florkowski:2017olj,Romatschke:2017vte,Spalinski:2017mel,Romatschke:2017acs,Behtash:2017wqg,Florkowski:2017jnz,Florkowski:2017ovw,Strickland:2017kux,Almaalol:2018ynx,Denicol:2018pak,Behtash:2018moe,Strickland:2018ayk,Heller:2018qvh,Behtash:2019qtk,Strickland:2019hff,Jaiswal:2019cju,Kurkela:2019set,Chattopadhyay:2019jqj,Brewer:2019oha, Attems:2016tby}. It has also been observed that many microscopic models, as well as various formulations of fluid dynamics, exhibit rapid information loss of some details (in particular the initial longitudinal pressure) of the initial condition leading to \emph{non-equilibrium attractor} behavior. The qualitative similarity of the attractors between different theories has been advocated to extend the applicability of the fluid-dynamic models to far-from-equilibrium regimes where the ordinary justification of fluid dynamics as a near-equilibrium expansion is questionable. 

Much less attention has been paid to the validity of the freeze-out procedure far from equilibrium. One of the challenges is that models with trivial momentum dependence, such as kinetic theory in relaxation time approximation, can give only limited information about the validity of the freeze-out procedure, whereas strongly coupled models without quasiparticle structure do not even possess underlying particle distributions. Here, we discuss the reconstruction of the particle distributions from the energy-momentum tensor in the Effective Kinetic Theory (EKT) for weak-coupling quantum chromodynamics (QCD) that becomes leading-order accurate in the limit of high center-of-mass energy collisions \cite{Arnold:2002zm}. The rich momentum-dependent structure of the EKT collision kernel allows for a non-trivial test of the freeze-out procedure in this theoretically clean limit. We follow 0+1d far-from-equilibrium Bjorken flow within this model and compare different moments of the distribution function to those predicted by hydrodynamic freeze-out prescriptions. We find that EKT seems to exhibit qualitatively similar far-from-equilibrium attractor behavior to RTA kinetic theory and Israel-Stewart-type hydrodynamics \cite{Heller:2015dha, Romatschke:2017vte, Kurkela:2019set} both at early (early-time or pullback attractor) and late times (late-time or hydrodynamic attractor).  We find that this attractive behavior is not restricted to the components of energy-momentum tensor but extends to other integral moments as well. We observe that the commonly used freeze-out prescriptions reproduce low-order moments of the distribution well at late times, however, they can fail at early times or when considering moments sensitive to momenta much larger than the temperature.  We discuss the implications for phenomenological fluid-dynamic modeling of small collision systems such as pp and pA. 

\vspace{2mm}
\noindent
{\em Methodology:} We make use of a numerical implementation of the effective kinetic theory (EKT) of Refs.~\cite{Arnold:2002zm,York:2014wja,Kurkela:2015qoa}. In parametrically isotropic systems,  EKT gives a leading order accurate description (in $\alpha_s$) of the QCD time evolution of the one-particle distribution function and allows for a numerical realization of the so-called bottom-up thermalization scenario \cite{Baier:2000sb}.  In practice, we solve the EKT Boltzmann equation for a gluonic plasma undergoing one-dimensional Bjorken expansion with transverse translational symmetry such that the effective Boltzmann equation reads \cite{Mueller:1999pi}
\be
-\frac{d f(\p)}{d\tau}+ \frac{p_z}{\tau}\partial_{p_z} f = \mathcal{C}_{1\leftrightarrow 2}[f(\p)] + \mathcal{C}_{2\leftrightarrow 2}[f(\p)] \, ,
\label{eq:ekt1}
\ee
where $f(\p)$ is the gluonic one-particle distribution function (per degree of freedom).  The elastic scattering term $\mathcal{C}_{2\leftrightarrow 2}$ and the effective inelastic term $\mathcal{C}_{1 \leftrightarrow 2}$ include physics of dynamical screening and Landau-Pomeranchuck-Migdal suppression and, in order to find the form of the collision kernels, self-energy and ladder resummations are required. For details,  we refer the reader to \cite{Arnold:2002zm,York:2014wja,Kurkela:2015qoa}. 

For the numerical solution of Eq.~\eqref{eq:ekt1}, we discretize $n(\p)= p^2  f(\p)$ on an optimized momentum-space grid and use Monte Carlo sampling to compute the integrals appearing in the elastic and inelastic collisional kernels. The algorithm used is based on Refs.~\cite{York:2014wja,PhysRevLett.115.182301} and exactly conserves energy while also exactly accounts for the particle number violation originating from the inelastic contributions to the collisional kernel. Due to the azimuthal symmetry of Bjorken flow,  one can discretize momentum-space on an effectively two-dimensional grid---here we use 250 $\times$ 2000 points in the $p$ and $x=\cos\theta$ directions, respectively.  The momenta $p$ are distributed on a logarithmic grid in the ranges $[0.02,45]\,\Lambda$, where $\Lambda$ is the typical energy scale of the initial condition. In all Figures presented herein, we used 't Hooft coupling $\lambda =N_c g^2 = 10$ corresponding to a specific shear viscosity of $\bar\eta = \eta/s \approx 0.62$ \cite{Arnold:2003zc,Keegan:2015avk}.

We follow the time evolution of a complete set of integral moments characterizing the momentum dependence of the distribution function~\cite{Strickland_2018}
\be
{\cal M}^{nm}(\tau) \equiv  \int \frac{d^3 p}{(2\pi)^3} \, p^{n-1} \, p_z^{2m} \, f(\tau,\p) \, ,
\label{eq:genmom1}
\ee
where $p = |{\bf p}|$.  Note that the energy density is given by $\varepsilon=\nu {\cal M}^{20}$, longitudinal pressure  by $P_L = \nu {\cal M}^{01}$, and number density by $n=\nu {\cal M}^{10}$ for $\nu$ degrees of freedom ($\nu = 2 d_A$ for $d_A$ adjoint colors of gluons).  The other moments do not have an interpretation in terms of the usual hydrodynamic moments considered in the literature, although the $m=0$ modes are simply related to the effective temperatures introduced in \cite{Kurkela:2018xxd,Kurkela:2018oqw}.

In our results, these moments will be scaled by their corresponding equilibrium values with \mbox{$\overline{\cal{M}}^{nm}(\tau) \equiv {\cal M}^{nm}(\tau) / {\cal M}^{nm}_{\rm eq}(\tau)$}, 
where, using a Bose distribution, one obtains
\be
{\cal M}^{nm}_{\rm eq} =  \frac{ T^{n+2m+2}\Gamma(n+2m+2) \zeta(n+2m+2)}{2 \pi^2 (2m+1)} \, .
\ee
The temperature $T$ here corresponds to the temperature of an equilibrium system with the same energy density, given by 
$T = (30 \varepsilon/\nu \pi^2)^{1/4}$.

The different moments are sensitive to different momentum regions of the distribution function and for future comparisons, we note that, in equilibrium, the typical momentum contributing to a given moment is  $\langle p \rangle^{nm}_{\rm eq} = {\cal M}^{n+1,m}_{\rm eq}/{\cal M}^{nm}_{\rm eq}$, giving, \emph{e.g.}, \mbox{$\langle p \rangle^{10}_{\rm eq} \simeq 2.7 \, T$},  $\langle p \rangle^{01}_{\rm eq} = \langle p \rangle^{20}_{\rm eq} \simeq 3.83 \, T$, $\langle p \rangle^{21}_{\rm eq} \simeq 5.95 \, T$, and $\langle p \rangle^{33}_{\rm eq} \simeq 11 \, T$.

%--------------------------------------------------------------------------------------------------------------------------------
\begin{figure*}[t!]
\includegraphics[width=\linewidth]{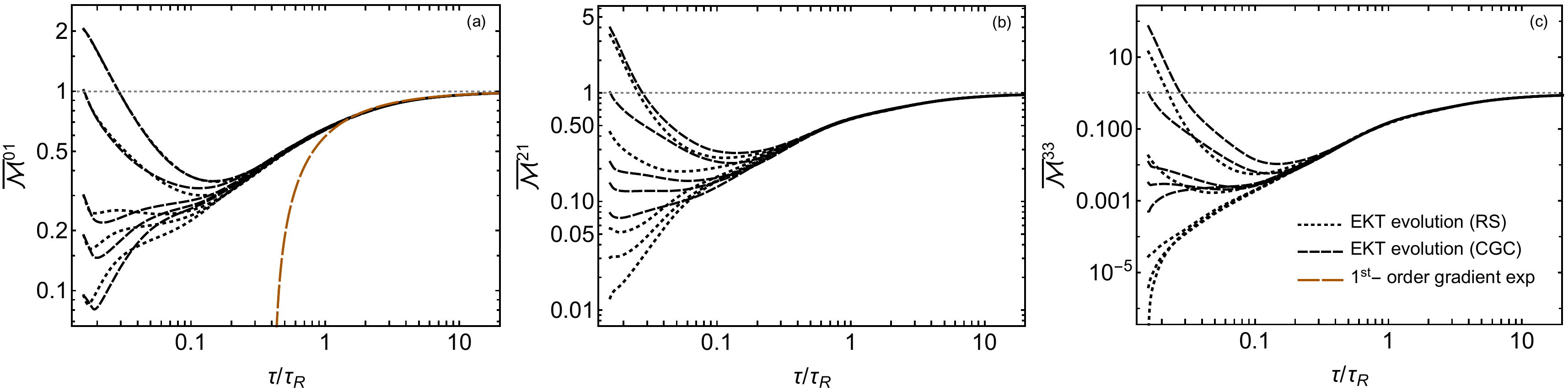}
\caption{Evolution of the scaled moments (a) $P_L/P_L^{\rm eq} = \overline{\cal{M}}^{01}$, (b) $\overline{\cal{M}}^{21}$, and (c) $\overline{\cal{M}}^{33}$  when varying the initial momentum-space anisotropy.  Black dotted and dashed lines show EKT evolution with RS and CGC initial conditions, respectively.  The orange long-dashed line shows the first-order gradient expansion result (Navier-Stokes). See supplemental Fig.~\ref{supfig1} for plots of more moments.
} 
\label{fig1}
\end{figure*}
%--------------------------------------------------------------------------------------------------------------------------------

\begin{figure*}[t!]
\includegraphics[width=\linewidth]{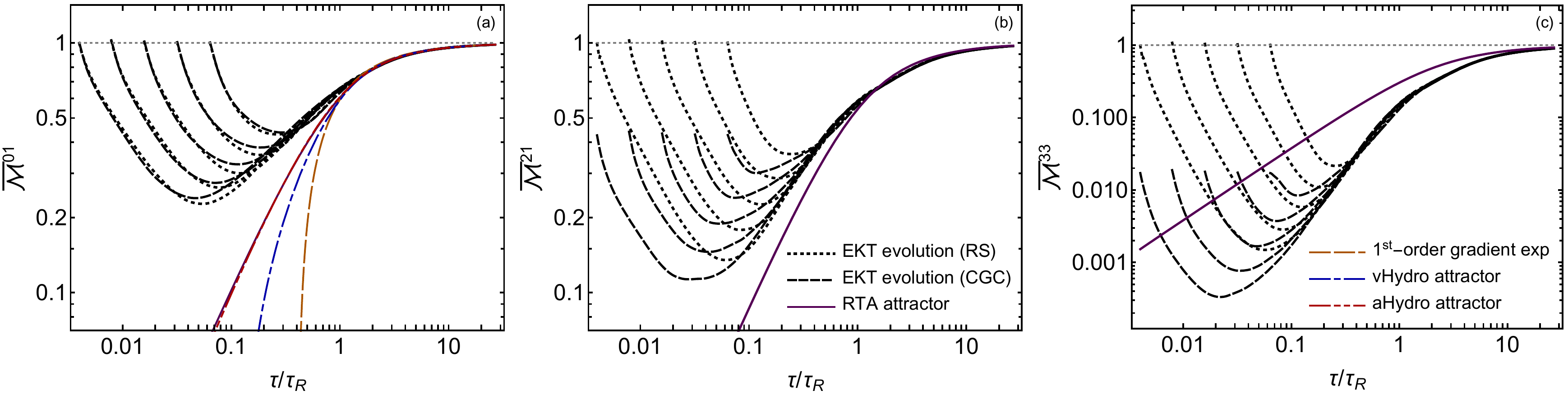}
\caption{Evolution of the scaled moments (a) $\overline{\cal{M}}^{01}$, (b) $\overline{\cal{M}}^{21}$, and (c) $\overline{\cal{M}}^{33}$ when varying the initialization time. Black dotted and dashed lines show EKT evolution with RS and CGC initial conditions, respectively.  The purple solid line is the exact RTA attractor, the orange long-dashed line is the first-order gradient expansion result, the blue dot-dashed line is the DNMR vHydro attractor, and the red dot-dot-dashed line is the aHydro attractor. See supplemental Fig.~\ref{supfig1} for plots of more moments.}
\label{fig2}
\end{figure*}

\vspace{2mm}
\noindent
{\em Results:}   In Figs.~\ref{fig1} and \ref{fig2} , we present results for the evolution of three scaled moments, $\overline{\cal{M}}^{01}$, $\overline{\cal{M}}^{21}$, and $\overline{\cal{M}}^{33}$, in panels (a), (b), and (c), respectively. These simulations have been initialized with either of the two following initial conditions:  (1) spheroidally-deformed thermal initial conditions which we will refer to as ``RS'' initial conditions \cite{Romatschke:2003ms} and (2) non-thermal color-glass-condenssate (CGC) inspired initial conditions \cite{Kurkela:2015qoa}.  In the first case, the initial gluonic one-particle distribution function is taken to be of the form
\be
f_{0,{\rm RS}}(\p) = f_{\rm Bose}\!\left(\sqrt{\p^2 + \xi_0 p_z^2 }/\Lambda_0\right) ,
\ee
where $-1 < \xi_0 < \infty$ encodes the initial momentum-anisotropy and $\Lambda_0$ is a temperature-like scale which sets the magnitude of the initial average transverse momentum.  In the second case, we take for the form of the initial gluonic one-particle distribution
\ba
f_{0,\rm CGC}(\p) &=&  \frac{2A}{\lambda} \frac{ \tilde \Lambda_0}{\sqrt{ \p^2 +\xi_0  p_z^2}} e^{-\frac{2}{3}\left(\p ^2 +  \xi_0 \hat{p}_z^2 \right)/\tilde \Lambda_0^2} \, .
\ea
This form has been used in several earlier works (see e.g.~\cite{Kurkela:2015qoa,Keegan:2016cpi,Kurkela:2018vqr,Kurkela:2018vqr,Kurkela:2018oqw,Kurkela:2018xxd}), and is motivated by the saturation framework, where the initial average transverse momentum scale $\tilde\Lambda_0$ is related to the saturation scale \mbox{$\tilde\Lambda_0 = \langle p_T \rangle_0 \approx 1.8\,Q_s$} \cite{Mueller_2000,Kovchegov_2001,Lappi:2011ju}.  The overall constant $A$ is set by fixing the initial energy density to match an expectation $\tau_0 \epsilon_0 = 0.358\, \nu Q_s^3/\lambda$ from a classical Yang-Mills simulation of Lappi~\cite{Lappi:2011ju}.

%--------------------------------------------------------------------------------------------------------------------------------
%--------------------------------------------------------------------------------------------------------------------------------

In both Figures, the dotted and dashed black lines correspond to EKT evolution using RS- and CGC-type initial conditions, respectively. Figure \ref{fig1} (a) shows the time evolution of the longitudinal pressure normalized by its equilibrium value. The different lines correspond to different initial $P_L/P_L^{\rm eq}= \overline{ \mathcal{M}}^{01}$ with momentum-space anisotropy parameters $\xi_0 \in \{-0.84,0,5.25,10.1,24\}$ at initial time $\tau_0 = 0.095\,(\nu/\varepsilon)^{1/4}$. 

The integral moments are plotted as a function of a rescaled time variable $\tau/\tau_R(\tau)$, which measures the age of the system in units of the instantaneous interaction time $\tau_R(\tau)$. As the density of the system changes, so does the interaction time scale, which is given by $\tau_R(\tau) = 4 \pi \bar\eta/T(\tau)$. Scaling time in this manner guarantees that, as long as the system is described by hydrodynamics close to thermal equilibrium, $\overline{ \mathcal{M}}^{01}$ will eventually be described by the first-order gradient expansion, $\overline{\mathcal{M}}^{01} = 1 - (120\zeta(5)/\pi^5) \tau_R/\tau$, at late times \cite{Heller:2016rtz,Strickland:2018ayk}. This is independent of the microscopic details or specific values of macroscopic hydrodynamic parameters.  This fluid-dynamic behavior is seen in Fig.~1~(a) where the evolutions of all the various different initial conditions eventually converge onto a universal curve -- the late-time attractor (see also \cite{Kurkela_2015,Giacalone:2019ldn}).  However, as observed also in simpler toy models, this collapse takes place before the system is well described by the hydrodynamic gradient expansion, the first order of which is shown in Figure \ref{fig1} (a) as an orange dashed line.  

While the late-time attractor behavior for the longitudinal pressure has been observed earlier in simplified kinetic theories, the solutions at hand allow us to study to what extent the attractive behavior determines the full overall shape of the distribution function.  Our first main finding is shown in panels (b) and (c) of Fig. 1 which display the time-evolution of two higher moments of the distribution function, $\overline{\mathcal{M}}^{21}$ and  $\overline{\mathcal{M}}^{33}$. We observe that the higher moments collapse to a universal curve on the same timescale as $\overline{\mathcal{M}}^{01}$, demonstrating that the universality extends beyond simple hydrodynamic moments and it is the entire distribution as a function of $\bf p$ that reaches an attractor form. For corresponding results for a large set of moments see the Supplemental Material associated with this paper. 

The timescale at which the different solutions to Eq.~\eqref{eq:ekt1} collapse in Figure 1 is roughly $\tau/\tau_R\sim 0.5$. While all theories must eventually collapse on a single curve, the time at which individual solutions collapse to the attractor depends on the details of the model. In \cite{Kurkela:2019set} two qualitatively different patterns were identified. In RTA kinetic theory and in IS hydrodynamics, the decay to the attractor took place by a powerlaw whose scale was set by the initial time $\tau_0$ such that a unique attractor exists at arbitrarily early time and can be found by studying initialisation with decreasing $\tau_0$. In contrast, in AdS/CFT, the decay to the attractor happens only at the time scale given by the decay of the quasi-normal modes. 

Figure 2 shows a set of solutions with  fixed initial conditions (RS or CGC) with successively decreasing $\tau_0$. The figure demonstrates that earlier initializations lead to faster decay to the attractor signifying $\tau_0$-scaling of the decay and the existence of an early-time (or pull-back \cite{Behtash:2019qtk}) attractor in EKT.   We note that at late times the attractor for both overoccupied (CGC) and the thermal (RS) initial conditions are the same.  This implies that upon reaching the attractor, the late-time evolution is not only insensitive to the initial longitudinal pressure of the initial condition but also to the initial occupancy and momentum profile for these physically motivated initial condition types and moderately large couplings ($\lambda=10$).  We note that in each run shown in Fig.~\ref{fig2} we observe a transition from purely free-streaming behavior to a collisionally-broadened longitudinal expansion related to the early stages of the bottom-up thermalization \cite{Baier:2000sb,Kurkela_2015} and to early-time  ``pre-scaling'' behavior seen in prior EKT studies~\cite{PhysRevLett.122.122301}. 

In Fig. 2 (a) we also compare the EKT attractor to other known attractors for $P_L/P^{\rm eq}_L$. The solid purple line corresponds to the exact solution for the attractor in kinetic theory in the RTA approximation \cite{Strickland:2018ayk,Florkowski:2013lza,Florkowski:2013lya,Baym:1984np,Baym:1985tna,Kurkela:2019set}, the orange long-dashed line is the first-order gradient expansion result, the blue dot-dashed line corresponds to a formulation of viscous fluid-dynamics used extensively in phenomenological description of heavy-ion collisions, namely second-order viscous hydrodynamics (vHydro) of Denicol et al \cite{Denicol:2010xn,Denicol:2011fa}, and the red dot-dot-dashed line corresponds to the anisotropic hydrodynamics (aHydro) attractor \cite{Strickland:2017kux,Florkowski:2010cf,Martinez:2010sc,Tinti:2013vba,Alqahtani:2017mhy}.  For details concerning how the attractors were determined in each case we refer the reader to the Supplemental Material.  We observe that while all of the attractors share some qualitatively similar features, the attractors of the different theories agree quantitatively only at $\tau/\tau_R \gg 1$ after the attractors follow the hydrodynamic gradient expansion. In particular, we emphasize that in vHydro the longitudinal pressure becomes negative at early times unlike in aHydro or EKT.

Panels (b) and (c) of Fig.~\ref{fig2} compare two higher-order moments of the RTA and EKT attractors. While the agreement between the two kinetic theories is rather good at late times for $n=0$ moments, the agreement becomes rapidly worse for increasing $n$~\footnote{\label{note1}We present table of results for a large set of moments in the supplemental material associated with this Letter (Figs.~\ref{supfig1} and \ref{supfig2}).}. This implies that while the $|{\bf p}|$-dependence of the collision kernel may be rather well approximated by the simple RTA kernel at these values of coupling $\lambda$, the simplified angular structure of the RTA does not fully capture the shape of the longitudinal structure of the distribution function.

%--------------------------------------------------------------------------------------------------------------------------------
\begin{figure*}[t!]
\includegraphics[width=\linewidth]{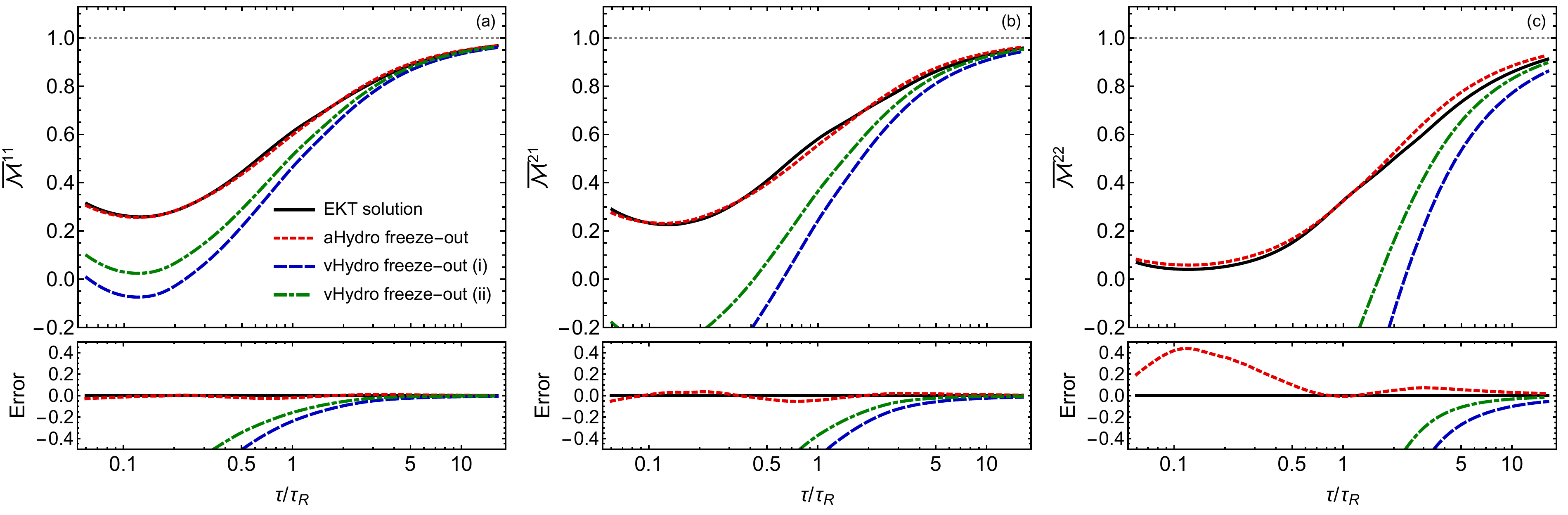}
\caption{Evolution of the scaled moments (a) $\overline{\cal{M}}^{11}$, (b) $\overline{\cal{M}}^{21}$, and (c) $\overline{\cal{M}}^{22}$.  The black solid line is a typical EKT evolution, the red-dashed line is the $P_L$-matched aHydro result for a given moment,  the blue and green dot-dashed lines are the corresponding vHydro results using Eqs.~\eqref{eq:fa} and \eqref{eq:fb}, respectively.  The relative error shown in the bottom panels is (${\rm approximation}/{\rm EKT} -1$). See supplemental Fig.~\ref{supfig2} for plots of more moments and comparison with our full set of runs.}
\label{fig3}
\end{figure*}
%--------------------------------------------------------------------------------------------------------------------------------

While the fluid-dynamic theories do not specify the higher moments of the distribution functions displayed in panels (b) and (c) of Fig.~2, it is a common practice to infer the full shape of the distribution from the shear components of the energy-momentum tensor only. 
For a given $T^{\mu\nu}$ the linearized viscous correction to the one-particle distribution function, $\delta f$ can be locally computed given an assumption of the collision kernel. Herein, we consider two possible forms for $\delta f$. The (i) \emph{quadratic ansatz}
\be
\frac{\delta f_{(i)}}{f_{\rm eq}(1+f_{\rm eq})} =  \frac{3 \overline\Pi}{16 T^2} ( p^2 - 3 p_z^2 ) \, ,
\label{eq:fa}
\ee
which results from a wide set of models including RTA with momentum-independent relaxation time, momentum diffusion approximation, scalar field theory, and from EKT in the leading-log approximation \cite{Dusling:2009df}. Here $\overline{\Pi} = \Pi/\epsilon = 1/3 - T^{zz}/\epsilon$ is shear viscous correction to the longitudinal pressure scaled by the energy density. At full leading-order, the EKT however has more structure; for large $p \gg T$, the EKT reduces to power law form of the (ii) \emph{LPM ansatz}
\be
\frac{\delta f_{(ii)}}{f_{\rm eq}(1+f_{\rm eq})} = \frac{16 \overline\Pi}{21 \sqrt{\pi}\, T^{3/2}} \! \left( p^{3/2} - \frac{3 p_z^2}{\sqrt{p}} \right) .
\label{eq:fb}
\ee
This $p^{1.5}$ power-law is numerically close to $\propto p^{1.38}$ that is found to describe the full EKT in the relevant momentum range in \cite{Dusling:2009df}.

Additionally, we consider a simple (iii) \emph{aHydro freeze-out ansatz} procedure that does not assume linearization around equilibrium. Instead, in the aHydro freeze-out ansatz, one assumes that the non-equilibrium distribution function can be approximated by a spheroidally-deformed Bose-distribution $f(p) = f_{\rm Bose}(\sqrt{{\bf p}^2 + \xi p_z^2}/\Lambda)$~\cite{Romatschke:2003ms,Florkowski:2010cf,Martinez:2010sc}.  To test this approach, we fix $\xi$ locally such that the energy-momentum tensor of the ansatz matches with that of the full simulation and then make predictions for higher-order moments.

The different moments obtained by the above prescriptions are compared to the EKT attractor solution in Fig.~3 (see the Supplemental Material for additional moments compared over our entire set of runs). At late times $\tau > 5\,\tau_R$, the low-order moments are described within a few percent by all the prescriptions, while some discrepancy remains even at $\tau  \sim 20\,\tau_R$ between the quadratic ansatz (i) and our EKT results. The agreement worsens gradually at earlier times and around \mbox{$\tau \sim \tau_R$} where the corrections to longitudinal pressure start to be sizable $P_L/P_L^{\rm eq} \sim 65\%$, ${\cal M}^{11}$ exhibits an approximately $20\%$ disagreement between EKT and both linearized ansatze. The disagreement increases for higher moments and for earlier times. In contrast, we observe rather good agreement between the aHydro ansatz and our EKT results at all times. 

\vspace{2mm}
\noindent
{\em Conclusions and Discussions:}
An important step in the phenomenological analysis of nuclear collisions is the freeze-out procedure in which the hydrodynamical fields are converted to particle distributions. In the current phenomenological practice, the quadratic ansatz (i) is widely used. This assumes linear deviations from thermal equilibrium, which is in stark contrast to the far-from-equilibrium conditions in which fluid-dynamical modelling is practiced in current phenomenological applications, in particular in modeling of small systems (see e.g. Refs.~\cite{Bozek:2013uha,Shen:2016zpp,Alqahtani:2016rth,Weller:2017tsr,Mantysaari:2017cni,Strickland:2018exs}).  To address whether these linearized procedures remain quantitatively predictive far from equilibrium, we have confronted them with far-from-equilibrium simulations of QCD effective kinetic theory.  Our results in Fig.~\ref{fig3} show that at least in this simplified framework -- admittedly quite far from the realities of phenomenological modeling -- the non-linear aHydro freeze-out ansatz performed better in reconstructing moments of the distribution function  compared to linearized ansatze in far-from-equilibrium systems.

To translate the rescaled time variable $\tau/\tau_R$ to physical units in real-world LHC nuclear collisions at \mbox{$\sqrt{s_{NN}} = 5.02$ TeV}, we set the scale of the simulations by using the fact that the initial entropy density with $\nu_{\rm eff} \simeq 40$ degrees of freedom satisfies \mbox{$\tau_0^{\rm hydro} s(\tau_0^{\rm hydro}) \simeq 7.62 \; {\rm GeV}^2$}, consistent with phenomenologically constrained values at LHC energies. For an initialization time of $\tau_0^{\rm hydro}=0.25$~fm/c, this corresponds to an initial temperature of $T_0^{\rm hydro} \approx 700$~MeV~\cite{Alqahtani:2017mhy}.  We also implicitly assume that the results as function of $\tau/\tau_R$ do not depend on $\bar\eta$ as seen for the hydrodynamic moments in \cite{Kurkela:2018vqr} and assume a value of $\bar \eta \approx 0.2$, which is consistent with the phenomenological extraction of the quantity.  

Using this setup and averaging over our full set of runs, the rescaled times \mbox{$\tau/\tau_R = \{0.2, 0.5, 1, 2, 5, 10\}$} map to \mbox{$ \tau \simeq \{0.32, 0.86,1.88,4.23,14.1,38.5\}$ fm/c}. This suggests that for $\tau_{\rm fo} \gg  5$ fm/c the lowest-order modes, which are sensitive to \mbox{$ p  \sim \textrm{ few }\,T$}, can be well described by both aHydro and linearized freeze-out prescriptions.  This implies that for central ultra-relativistic heavy-ion collisions one can have a faithful reproduction of the low-momentum part of the freeze-out distribution function.  However, when considering higher-moments, which are sensitive to higher momenta $\langle p \rangle^{nm}$, or applying early-time freeze-out for smaller systems such as peripheral nucleus-nucleus collisions and proton-nucleus collision, the aHydro freeze-out ansatz is favored.

In closing, we note that the current study was performed in a very simple setting with one-dimensional Bjorken flow and considering only massless gluonic degrees of freedom.  We leave, for the future, extensions to more realistic geometries \cite{Romatschke:2017acs,Behtash:2017wqg,Denicol:2018pak} and inclusion of quark degrees of freedom \cite{Kurkela:2018xxd,Kurkela:2018oqw}. 

%\vspace{2mm}
\acknowledgements{{\it Acknowledgements:}
We thank A. Mazeliauskas for useful discussions.  We also thank the Ohio Supercomputer Center under the auspices of Project No.~PGS0251.  M.S. and D.A. were supported by the U.S. Department of Energy, Office of Science, Office of Nuclear Physics Award No.~DE-SC0013470. 

%%%%%%%%%%%%%%%%%%%%%%%%%%%%%%%%%%%%%%%%%%%%%%%%%%%%%%%%%%%%%%
\bibliography{pseudo}
%%%%%%%%%%%%%%%%%%%%%%%%%%%%%%%%%%%%%%%%%%%%%%%%%%%%%%%%%%%%%%

%\clearpage 
\widetext{\ }

%%%%%%%%%%%%%%%%%%%%%%%%%%%%%%%%%%%%%%%%%%%%%%%%%%%%%%%%%%%%%%
\section{Supplemental material}
\label{supp}
%%%%%%%%%%%%%%%%%%%%%%%%%%%%%%%%%%%%%%%%%%%%%%%%%%%%%%%%%%%%%%

Here we provide some brief supplemental material divided into two subsections.  In the first part, we present results for a large set of scaled moments.  In the second part, we provide the details concerning each of the various attractors we compare to in the main body and provide references to the appropriate literature for each.  

%%%%%%%%%%%%%%%%%%%%%%%%%%%%%%%%%%%%%%%%%%%%%%%%%%%%%%%%%%%%%%
\subsection{1. Higher-order moments}
%%%%%%%%%%%%%%%%%%%%%%%%%%%%%%%%%%%%%%%%%%%%%%%%%%%%%%%%%%%%%%

As additional information, in Fig.~\ref{supfig1} we plot the evolution of scaled moments with $\overline{\cal{M}}^{nm}$ with $n \in \{0,1,2\}$ and $m \in \{0,1,2,3\}$.  As can be seen from Fig.~\ref{supfig1}, we find evidence of a non-equilibrium attractor both at early and late times in all moments. In Fig.~\ref{supfig2} we plot all EKT runs compared to the three freeze-out prescriptions -- quadratic ansatz (i), LPM ansatz (ii), and aHydro freeze-out ansatz (iii) -- with $n \in \{0,1,2\}$ and $m \in \{0,1,2,3\}$.  As can be seen from Fig.~\ref{supfig2}, the aHydro freeze-out prescription (iii) performs well at all times and at late times the LPM ansatz (ii) is superior to the quadratic ansatz (i).  At early times and in high-order moments, we see that both linearized schemes (i) and (ii) predict negative values for the moments.

%-----------------------------------------------------------------------------------
\begin{figure}[t!]
\includegraphics[width=1\linewidth]{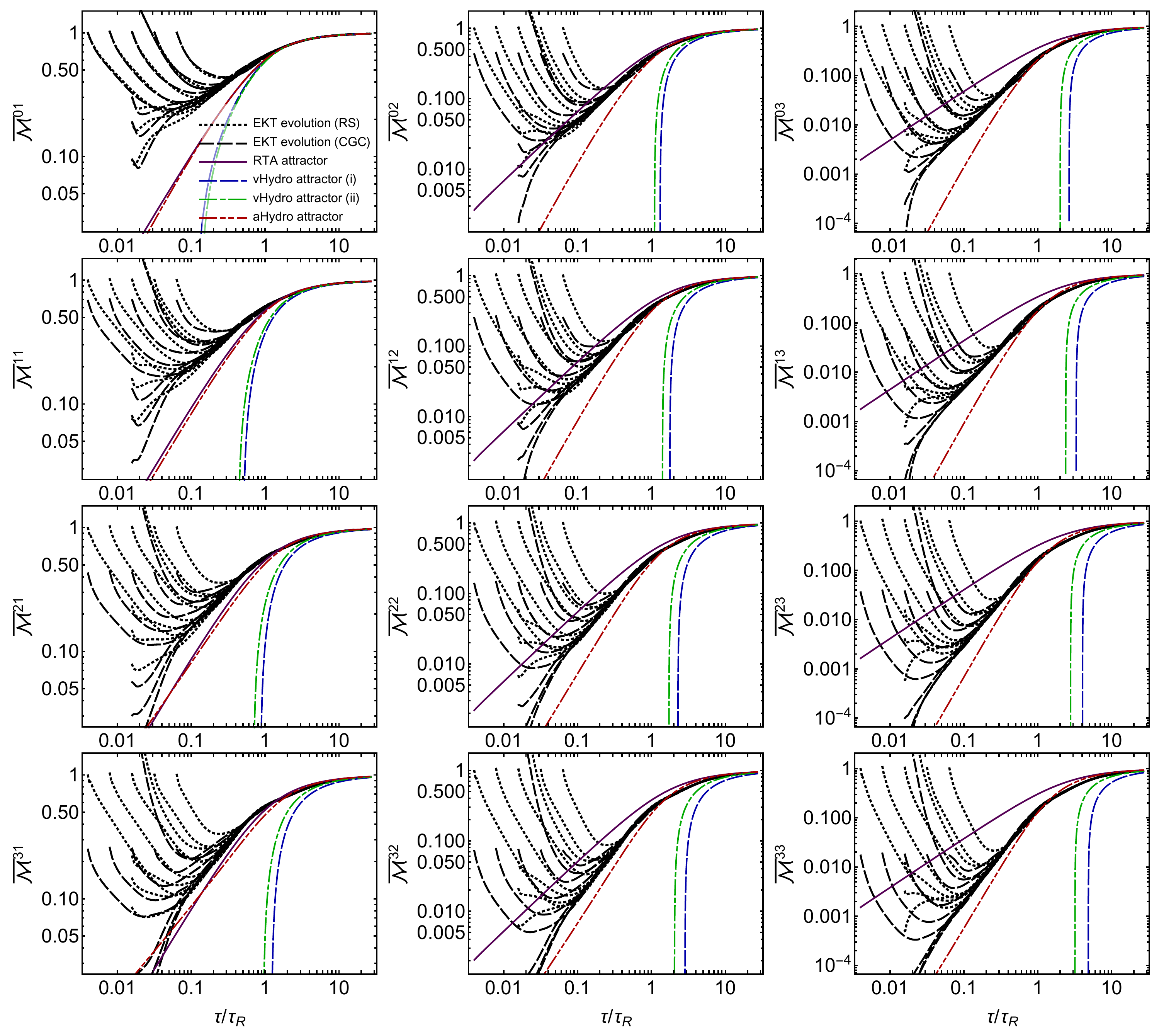}
\caption{Evolution of the scaled moments $\overline{\cal{M}}^{nm}$ with $n \in \{0,1,2\}$ and $m \in \{0,1,2,3\}$.  Black dotted and dashed lines show EKT evolution with RS amd CGC initial conditions, respectively.  The purple solid line is the exact RTA attractor, the blue long-dashed line is the DNMR vHydro attractor using $\delta f$ parameterization (i), the green dot-dashed line is the DNMR vHydro attractor using $\delta f$ parameterization (ii), and the red dot-dot-dashed line is the aHydro attractor.}
\label{supfig1}
\end{figure}
%-----------------------------------------------------------------------------------

%-----------------------------------------------------------------------------------
\begin{figure}[t!]
\includegraphics[width=1\linewidth]{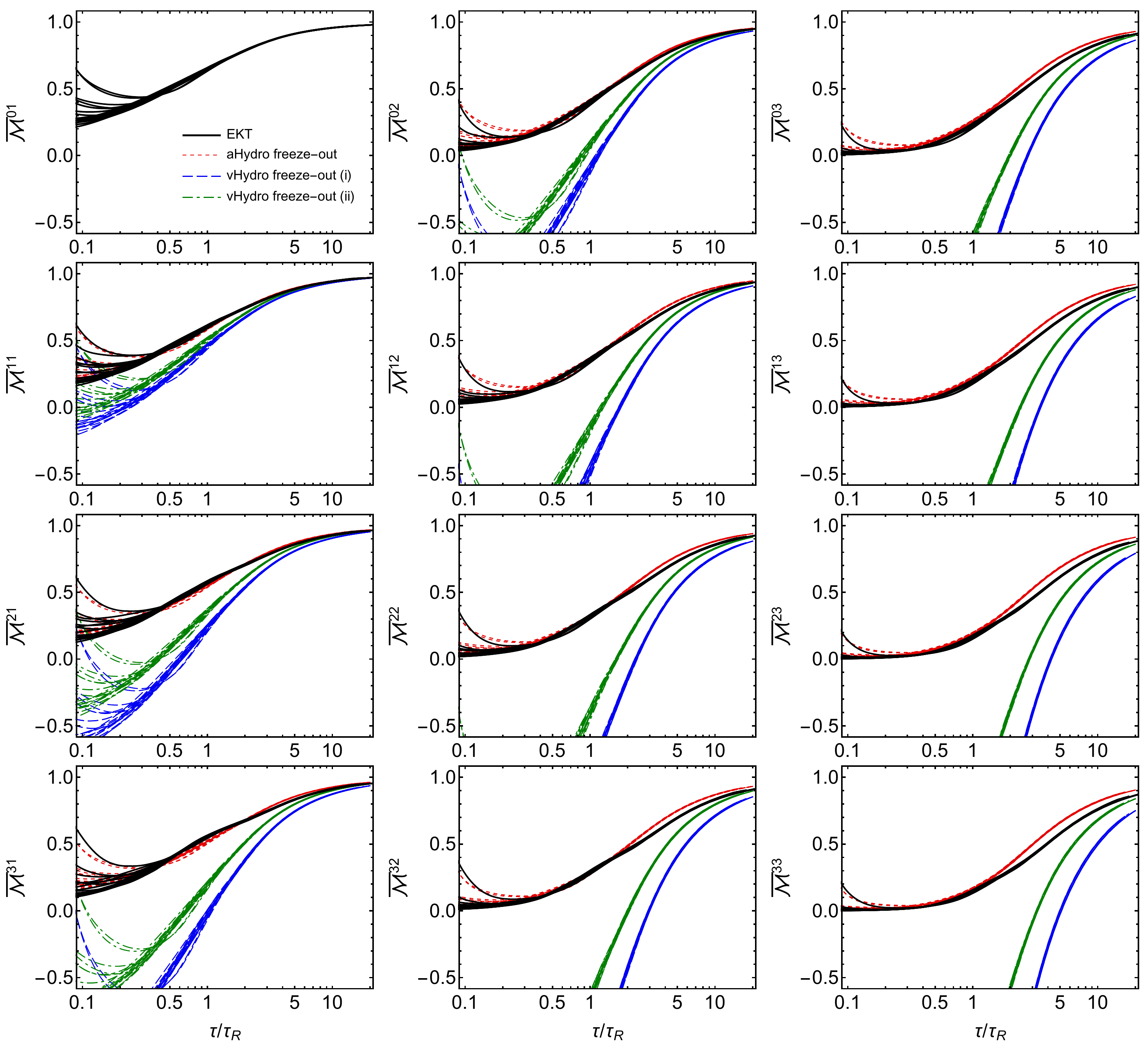}
\caption{Evolution of the scaled moments $\overline{\cal{M}}^{nm}$ with $n \in \{0,1,2\}$ and $m \in \{0,1,2,3\}$ for all runs (CGC +RS) shown as black lines. Other lines -- red-dashed, blue long-dashed, green dot-dashed -- correspond to results obtained in three freeze-out scenarios, aHydro, vHydro (i), and vHydro (ii), respectively. Note that all scenarios agree on $\overline{\cal M}^{01}$ by construction ($P_L$-matching). The other panels are predictions based on this matching.}
\label{supfig2}
\end{figure}
%-----------------------------------------------------------------------------------

%%%%%%%%%%%%%%%%%%%%%%%%%%%%%%%%%%%%%%%%%%%%%%%%%%%%%%%%%%%%%%
\subsection{2. RTA, vHydro, and aHydro attractors and freeze-out prescriptions}
\label{supp:attractors}
%%%%%%%%%%%%%%%%%%%%%%%%%%%%%%%%%%%%%%%%%%%%%%%%%%%%%%%%%%%%%%

It has been established that RTA in a one-dimensionally Bjorken expanding geometry has an attractor both at early and late times \cite{Romatschke:2017vte,Strickland:2018ayk,Kurkela:2019set}.
In the main body of this paper, we compared our EKT numerical results to (1) moments obtained from RTA using the attractors emerging from the exact solution to the conformal 0+1d RTA kinetic equation \cite{Strickland:2018ayk,Florkowski:2013lza,Florkowski:2013lya,Baym:1984np,Baym:1985tna,Kurkela:2019set}, (2) second-order viscous hydrodynamics (vHydro) \cite{Teaney:2009qa,Dusling:2009df}, and (3) anisotropic hydrodynamics \cite{Florkowski:2010cf,Martinez:2010sc,Alqahtani:2017mhy}.  

For our comparisons with second-order vHydro, we determine the attractor solution of the 0+1d conformal vHydro equations of Denicol et al (DNMR) \cite{Denicol:2010xn,Denicol:2011fa} by numerically solving two coupled ordinary differential equations
\be
\tau \frac{d\log \epsilon}{d\tau}  = -\frac{4}{3} + \overline\Pi \, ,
\label{eq:firstmom}
\ee 
with $\Pi = P_{\rm eq} - P_{\rm L}$ being the shear viscous correction to the longitudinal pressure, $\overline\Pi = \Pi/\epsilon$, and
\be
\frac{d\Pi}{d\tau} = \frac{4\eta}{3\tau\tau_\pi} - \frac{38}{21}\frac{\Pi}{\tau} - \frac{\Pi}{\tau_\pi} \, ,
\label{2ndorderhydro}
\ee
where $\eta$ is the shear viscosity and $\tau_\pi = 5.1 \bar\eta/T$ \cite{York:2008rr}.  To obtain the DNMR vHydro attractor, we use the boundary condition $\overline\Pi(0^+) = (\sqrt{464865}-255)/1091$~\cite{Strickland:2017kux,Heller:2015dha}.   The resulting attractor solution for $\overline\Pi$ is then used to compute the viscous correction to the one-particle distribution function, $\delta f$. 

In the main body of the Letter we considered two possible forms for the linearized $\delta f$, given by Eqs.~\eqref{eq:fa} and \eqref{eq:fb}, corresponding to cases (i) and (ii), respectively. Using these forms, one can compute the vHydro predictions for the evolution of all scaled moments.  In case (i), one obtains
\be
\overline{\cal M}^{nm}_{\rm vHydro,(i)} = 1 - \frac{3 m (n+2m+2)(n+2m+3)\zeta(n+2m+3)}{4(2m+3)\zeta(n+2m+2)} \overline\Pi \, , \;
\label{eq:vhydromomsa}
\ee
and, in case (ii), one obtains
\be
\overline{\cal M}^{nm}_{\rm vHydro,(ii)} = 1 - \frac{64  m \Gamma\!\left(n+2m+\frac{7}{2}\right)\zeta\!\left(n+2m+\frac{5}{2}\right)}{21 \sqrt{\pi} (2 m+3) \Gamma(n+2m+2)\zeta(n+2m+2)} \overline\Pi \, . \;
\label{eq:vhydromomsb}
\ee

For the case of anisotropic hydrodynamics, we follow Ref.~\cite{Strickland:2017kux}.  For a conformal 0+1d system only one anisotropy parameter, $\xi$, is required such that $f(p) = f_{\rm Bose}(\sqrt{{\bf p}^2 + \xi p_z^2}/\Lambda)$.  The $\xi$ equation of motion is obtained from a linear combination of second-moments of the Boltzmann equation \cite{Tinti:2013vba}.  The $\xi$ equation can be combined with the requirement of energy-momentum conservation to obtain a closed set of evolution equations which include the scale $\Lambda$.  For a 0+1d system undergoing Bjorken expansion, the energy conservation equation is the same as in vHydro \eqref{eq:firstmom} but with
\be
\overline\Pi = \frac{1}{3} \left[ 1 - \frac{{\cal R}_L(\xi)}{\cal R(\xi)} \right] ,
\label{eq:pixirel}
\ee
with $2{\cal R}(x) = (1+x)^{-1} + x^{-1/2}\arctan\sqrt{x}$ and ${\cal R}_L(x) = 3(x+1)[(1+x) {\cal R}(x) -1]/x$.
Assuming an RTA collisional-kernel, the second moment equation reduces to
\be
\frac{1}{1+\xi} \dot\xi - \frac{2}{\tau} + \frac{{\cal R}^{5/4}(\xi)}{\tau_{\rm eq}} \xi \sqrt{1+\xi} = 0\, ,
\label{eq:2ndmomf}
\ee
with $\tau_{\rm eq} = 5.1\bar\eta/T$ and $T = {\cal R}^{1/4}(\xi) \Lambda$.  To obtain the aHydro attractor, one solves Eqs.~\eqref{eq:firstmom}, \eqref{eq:pixirel}, and \eqref{eq:2ndmomf} numerically with the boundary condition $\xi(0^+) = \infty$ \cite{Strickland:2017kux}.  Once the time evolution of the anisotropy parameter is obtained, the aHydro scaled moments can be computed using \cite{Strickland:2018ayk}
\be
\overline{\cal M}^{nm}_{\rm aHydro} = 2^{(n+2m-2)/4} \frac{(2m+1) {\cal H}^{nm}(\alpha)}{[{\cal H}^{20}(\alpha)]^{(n+2m+2)/4}} \, ,
\label{eq:ahydromoms}
\ee
where $\alpha=(1+\xi)^{-1/2}$ and
$
{\cal H}^{nm}(y) = \tfrac{2y^{2m+1}}{2m+1}  {}_2F_1(\tfrac{1}{2}+m,\tfrac{1-n}{2};\tfrac{3}{2}+m;1-y^2)  \, .
$

\end{document}